\documentclass[12pt,preprint]{aastex}
\newcommand{\n}{\ensuremath{\mem{n}}}

\newcommand{\p}{\ensuremath{\mem{p}}}

\newcommand{\hevi}{\ensuremath{^{4}\mem{He}}}

\newcommand{\cdr}{\ensuremath{^{13}\mem{C}}}
\newcommand{\czw}{\ensuremath{^{12}\mem{C}}}

\newcommand{\ndr}{\ensuremath{^{13}\mem{N}}}
\newcommand{\nvi}{\ensuremath{^{14}\mem{N}}}
\newcommand{\nfu}{\ensuremath{^{15}\mem{N}}}

\newcommand{\ose}{\ensuremath{^{16}\mem{O}}}

\newcommand{\oac}{\ensuremath{^{18}\mem{O}}}

\newcommand{\fne}{\ensuremath{^{19}\mem{F}}}

\newcommand{\nezw}{\ensuremath{^{22}\mem{Ne}}}

\newcommand{\msun}{\ensuremath{\, {\rm M}_\odot}}

\newcommand{\kelv}{\ensuremath{\,\mathrm K}}
\newcommand{\jahre}{\ensuremath{\, \mathrm{yr}}}
\newcommand{\mdot}{\ensuremath{\dot{M}}}
\newcommand{\beq}{\begin{equation}}
\newcommand{\beqa}{\begin{eqnarray}}
\newcommand{\eeq}{\end{equation}}
\newcommand{\eeqa}{\end{eqnarray}}
\newcommand{\bedis}{\begin{displaymath}}
\newcommand{\edis}{\end{displaymath}}

\newcommand{\mem}[1]{\ensuremath{\mathrm{ #1}}}

\newcommand{\abb}[1]{Fig.\,\ref{#1}}

\usepackage{natbib}
\def\degree{\ifmmode {^\circ}\else {$^\circ$}\fi}
\def\rstar{\ifmmode {\, R_{\star}}\else $R_{\star}$\fi}
\def\msol{\ifmmode {\, M_{\odot}}\else $M_{\odot}$\fi}
\def\rsol{\ifmmode {\, R_{\odot}}\else $R_{\odot}$\fi}
\def\lsol{\ifmmode {\, L_{\odot}}\else $L_{\odot}$\fi}
\def\msolyr{\ifmmode {\, M_{\odot}\,{\rm yr}^{-1}}\else $M_{\odot}\,{\rm yr}^{-1}$\fi}
\def\mdot{\ifmmode {\,\dot{M}}\else $\dot{M}$\fi}
\def\mdotyr{\ifmmode {\,\dot{M}\,yr^{-1}}\else $\dot{M}\,yr^{-1}$\fi}

\begin{document}

\title{Very Large Excesses of $^{18}$O in Hydrogen-Deficient
Carbon and R Coronae Borealis Stars: Evidence for White Dwarf Mergers}

\author{ Geoffrey C. Clayton\altaffilmark{1}, T.R. Geballe\altaffilmark{2}, Falk
Herwig\altaffilmark{3,4},  Christopher Fryer\altaffilmark{4}, and Martin 
Asplund\altaffilmark{5}}

\altaffiltext{1}{Department of Physics \& Astronomy, Louisiana State
University, Baton Rouge, LA 70803; gclayton@fenway.phys.lsu.edu}

\altaffiltext{2}{Gemini Observatory, 670 N. A'ohoku Place, Hilo, HI 96720;  
tgeballe@gemini.edu}

\altaffiltext{3}{Keele Astrophysics Group, School of Physical and 
Geographical Sciences, Keele University, Staffordshire ST5 5BG, UK; fherwig@astro.keele.ac.uk}

\altaffiltext{4}{Los Alamos National Laboratory, Los Alamos, NM 87545;
clfreyer@lanl.gov}

\altaffiltext{5}{Research School of Astronomy and Astrophysics, Mount 
Stromlo Observatory, Cotter Road, Weston, ACT 2611, Australia; 
martin@mso.anu.edu.au}

%\altaffiltext{6}{Department of Chemistry, University of Arizona P.O. Box 210041, 1306 E. University Blvd., Tucson, Arizona, 85721-0041; emilyt@as.arizona.edu}

\begin{abstract}

We have found that at least seven hydrogen-deficient carbon (HdC) and
R Coronae Borealis (RCB) stars, have $^{16}$O/$^{18}$O ratios close to
and in some cases less than unity, values that are orders of magnitude
lower than measured in other stars (the Solar value is 500). Greatly
enhanced $^{18}$O is evident in every HdC and RCB we have measured
that is cool enough to have detectable CO bands. The three HdC stars measured have $^{16}$O/$^{18}$O~$<$~1, lower values than any of the RCB stars. These discoveries are important clues in determining the evolutionary
pathways of HdC and RCB stars, for which two models have
been proposed: the double degenerate (white dwarf (WD) merger), and the
final helium-shell flash (FF). No overproduction of $^{18}$O is expected in the FF scenario. We have quantitatively explored the idea that HdC and
RCB stars originate in the mergers of CO- and He-WDs. The merger process is estimated to take only a few days, with accretion rates of $150\msun~yr^{-1}$ producing temperatures at the base of the accreted envelope of $1.2 - 1.9 \times 10^{8}\kelv$. Analysis of a simplified one-zone calculation shows that nucleosynthesis in the dynamically accreting material may provide a suitable environment for a significant production of $^{18}$O, leading
to very low values of $^{16}$O/$^{18}$O, similar to those observed. We also find
qualitative agreement with observed values of $^{12}$C/$^{13}$C and with the CNO elemental ratios. H-admixture during the accretion process from the small H-rich C/O WD envelope may play an important role in producing the observed
abundances. Overall our analysis shows that WD mergers may very well
be the progenitors of O$^{18}$-rich RCB and HdC stars, and that more
detailed simulations and modeling are justified.

\end{abstract}

% Keywords should be included, but they are not printed in the hardcopy.


\section{Introduction}

Among the hydrogen-deficient post-asymptotic giant branch (post-AGB) stars
are the R Coronae Borealis (RCB) stars, a small group of carbon-rich
supergiants.  About 50 RCB stars are known in the Galaxy and the
Magellanic Clouds
\citep{2005ApJ...631L.147K,2005AJ....130.2293Z,2001ApJ...554..298A,1996PASP..108..225C}.  
Their defining characteristics are hydrogen deficiency and unusual variability - RCB stars
undergo massive declines of up to 8 mag due to the formation of carbon
dust at irregular intervals. Apparently related to the RCB stars are the
hydrogen-deficient carbon (HdC) stars.  The five known HdC stars are
similar to the RCB stars spectroscopically but do not show declines or IR
excesses \citep{1967MNRAS.137..119W}.

Two scenarios have been proposed for the origin of an RCB star: the double
degenerate (DD) and the final helium-shell flash (FF) models
\citep{1996ApJ...456..750I,2002MNRAS.333..121S}. The former involves the
merger of a CO- and a He-white dwarf (WD) \citep{1984ApJ...277..355W}. In the latter, thought to occur
in 20\% of all AGB stars, a star evolving into a planetary nebula (PN)
central star is blown up to supergiant size by a FF \citep{1977PASJ...29..331F,1979sss..meet..155R}. Three stars
(Sakurai's Object, V605 Aql, and FG Sge) have been observed to undergo FF
outbursts that transformed them from hot evolved PN central stars into cool
giants with spectral properties similar to RCB stars
\citep{1998A&A...332..651A,1999A&A...343..507A, 2000A&A...353..287A,1997AJ....114.2679C,1998ApJS..114..133G}. In 
two of these, Sakurai's Object and V605 Aql, the FFs are thought to be the
result of Very Late Thermal Pulses (VLTPs)\footnote{FG Sge is thought to be the result of a Late Thermal Pulse (LTP) and is not considered further here \citep{2001ApJ...554L..71H}.}, FFs that occur on the WD
cooling track \citep[e.g.,][]{2001ApJ...554L..71H}, because their post-outburst evolutions have been so rapid and because in some cases an ionized nebula was found surrounding the cool post-outburst object \citep{1971ApJ...170..547F,1996ApJ...468L.111D}. V605 Aql has evolved from T$_{eff}$=5000 K in 1921 (two years after its outburst) to
95,000 K today \citep{2006ApJ...646L..69C} and now has abundances similar
to those seen in Wolf-Rayet [WC] central stars of PNe, with ~55\% He, and
40\% C, and not similar to those seen in the RCB and HdC stars. Sakurai's Object became enshrouded in dust three years after its outburst, and thus most of its current stellar properties cannot be determined. However, just after their outbursts, both V605 Aql and Sakurai's Object were almost indistinguishable from the RCB stars, in abundances (except for the presence of $^{13}$C), temperature, and absolute luminosity. Nevertheless, the very short observed timescales for this RCB-like phase 
make it unlikely that objects such as these can account for even the small number of RCB stars known in the Galaxy \citep{2006ApJ...646L..69C}.

Recently, \citet{2005ApJ...623L.141C} discovered that $^{16}$O/$^{18}$O$~\lesssim$~1 in the HdC star, HD 137613, from a spectrum of the first overtone bands of CO at 2.3-2.4 \micron. As the bands of 
$^{12}$C$^{16}$O are of typical strength for a cool star, the measured ratio implies a huge enhancement of $^{18}$O rather than a depletion of $^{16}$O.  The isotopic ratio is orders of magnitude less than that measured in other stars \citep{2005ApJ...623L.141C}. The value of $^{16}$O/$^{18}$O is $\sim$500 in the solar neighborhood
\citep{2006A&A...456..675S,2002ApJ...578..862G} and varies from 200 to 600 in the Galactic interstellar medium \citep{1994ARA&A..32..191W}. Only one other star, the post-AGB star HR 4049, has been found to have highly enhanced $^{18}$O
\citep{2001A&A...367L...1C}. HR 4049 is a binary and its enhanced oxygen
isotopes ($^{16}$O/$^{18}$O~$\approx~7$ and $^{16}$O/$^{17}$O~$\approx~8$) are found in the circumbinary disk material.  High abundances of $^{18}$O have also been measured in pre-solar graphite grains in the Murchison meteorite
\citep{1995ApJ...447L.147A} and are attributed to material processed in
massive Wolf-Rayet stars. In both of these other cases (HR 4049 and
meteorites), the abundance of $^{18}$O relative to $^{16}$O is several
times less than in HD 137613.

In this paper, we analyze new high S/N spectra of the K-band CO bands in
HD~137613 and nine additional RCB and HdC stars to investigate $^{16}$O/$^{18}$O
in these stars. The sample includes all known HdC stars and five RCB stars cool enough to have prominent CO bands. \citet{2005ApJ...623L.141C}  discussed the failure of several nucleosynthesis scenarios to explain the unusually small
$^{16}$O/$^{18}$O ratio in HD~137613. In the light of the new observations, we explore the idea that HdC and RCB stars originate in DD mergers of a CO and a He-WD. We first investigate whether the conditions exist in a WD merger for nucleosynthesis to take place and then estimate the results of the nucleosynthesis.

\section{Observations and Data Reduction}

We obtained a more extensive K-band spectrum of HD 137613 than the one
published by \citet{2005ApJ...623L.141C} on UT 2005 March 04 at the 3.8~m United
Kingdom Infrared Telescope (UKIRT).  The new observation used the
facility instrument UIST with its long K grism and 0\farcs24 (2
pixel-wide) slit to cover 2.20-2.51 \micron~at a resolving power of
$\sim$3500 \citep{2004SPIE.5492.1160R}.
Standard observing and reduction techniques were employed.  A
wavelength calibration accuracy of better than 0.0001 $\mu$m was achieved
using the spectrum of an argon arc lamp. The calibration star used to
remove telluric absorption lines and flux-calibrate the spectrum of HD
137613 was HR 5514 (B9V). The reduced spectrum is plotted in Fig. 1, with
vertical lines indicating the bandhead wavelengths. It shows, in addition
to the 2-0, 3-1, and 4-2 bands of $^{12}$C$^{18}$O discovered by
\citet{2005ApJ...623L.141C}, the 5-3, 6-4, and 7-5 bands. In addition to
the six bands of $^{12}$C$^{18}$O, at least seven and probably eight bands
of $^{12}$C$^{16}$O are seen. As in the previously published K-band
spectrum of HD 137613 the strengths of the $^{12}$C$^{18}$O and
$^{12}$C$^{16}$O bands are approximately equal, implying near equal
abundances of the two isotopomers. The spectrum shows no evidence for
either $^{13}$C$^{16}$O or $^{12}$C$^{17}$O.

The Gemini-S facility spectrometer, GNIRS was employed on UT 2005
September 26-28 to obtain spectra of nine HdC and RCB stars. The
spectrograph was used in the short camera mode and with the 0\farcs3 (2
pixel-wide) slit to cover 2.27 to 2.51 \micron~at R=5900.  The observations were reduced by dividing the spectrum of each RCB and HdC star by that of a nearby bright early A-type dwarf star to remove the instrumental response and the effects of atmospheric absorption lines. Wavelength calibration was obtained from these telluric lines and is accurate to better than 0.0001~$\mu$m.

The complete observing sample is listed in Table 1 and the
2.28--2.45~$\mu$m portions of the spectra are plotted in Fig. 2. 
Also included is the previously published spectrum of the RCB star, Z UMi \citep{2005AJ....130..256T}. 
The
wavelengths of the bandheads of $^{12}$C$^{16}$O and $^{12}$C$^{18}$O are
indicated by vertical lines. The signal-to-noise ratios of the spectra are
100 or more. The spectra have been shifted to correct for the stars'
measured heliocentric radial velocities
\citep{1997MNRAS.285..266L,1995A&AS..114..269D,1997PASP..109..796G}. One star, ES Aql, does
not have a published radial velocity. We have estimated it to be $\sim$55
km s$^{-1}$ and made the correction in Fig. 2.

\section{Preliminary Results}

Seven of the eleven stars in Fig. 2 show, in addition to prominent bands
of $^{12}$C$^{16}$O, compelling evidence for bands of $^{12}$C$^{18}$O.
These are the HdC stars HD~137613, HD~175893, and HD~182040, and the RCB
stars WX CrA, ES Aql, S Aps, and SV Sge.  One additional RCB star that we observed, U Aqr, shows marginal evidence for $^{12}$C$^{18}$O. We also find marginal evidence for $^{12}$C$^{18}$O in the the spectrum of the RCB star, Z UMi \citep{2005AJ....130..256T}.

In the spectra of three of the HdC stars, HD~137613 and HD~175893, and
HD~182040, and in the spectrum of one RCB star, WX~CrA, the
$^{12}$C$^{18}$O bands are about as strong or stronger than the
$^{12}$C$^{16}$O bands.  
The most remarkable case is the HdC star
HD~175893, for which the maximum depths of the bands of $^{12}$C$^{18}$O
are more than twice as deep as those of $^{12}$C$^{16}$O.  The least
secure of these four cases is the HdC star HD~182040, because the bands
are weak, but there is accurate wavelength correspondence between observed absorption maxima and their expected wavelengths. 
As in the case of HD~175893, the $^{12}$C$^{18}$O bands in its spectrum are noticeably stronger than the $^{12}$C$^{16}$O bands. For the third HdC star, HD~137613, the $^{12}$C$^{18}$O bands appear to be slightly stronger than the $^{12}$C$^{16}$O bands, whereas in the RCB star with the strongest $^{12}$C$^{18}$O, WX CrA, the band strengths do not exceed those of $^{12}$C$^{16}$O. 
In three other RCB stars, ES~Aql, S~Aps, and SV~Sge, the $^{12}$C$^{18}$O bands are prominent, but are considerably weaker than than the $^{12}$C$^{16}$O bands.
Two HdC stars, HD~148839 and HD~173409, have no significant CO bands,
presumably because their relatively high photospheric temperatures preclude the
formation of detectable amounts of CO.  Thus for them no determinations of the relative abundances of the CO isotopomers are possible.

For the sources with weaker absorptions (U Aqr, Z UMi, and HD 182040), there is
some possibility that the features at the expected wavelengths of the
$^{12}$C$^{18}$O bandheads are due to some other species (e.g., CN).
However, the accurate wavelength agreement of the
$^{12}$C$^{18}$O bandheads (in terms both of absolute wavelength and wavelength difference from nearby $^{12}$C$^{16}$O bandheads), plus the fact that the
absorptions at those wavelengths are among the strongest (apart from the
$^{12}$C$^{16}$O absorptions) in each spectrum gives us confidence,
especially in the case of HD~182040, that we are detecting
$^{12}$C$^{18}$O. None of the spectra show evidence for absorption bands of $^{13}$C$^{16}$O, which are present in the spectra of some carbon stars. Very high $^{12}$C/$^{13}$C ratios are one of the trademarks of the RCB and HdC stars \citep{2001ApJ...554..298A,1996PASP..108..225C,1967MNRAS.137..119W}.

\section{Estimates of $^{16}$O/$^{18}$O}

To further confirm the identification of the $^{12}$C$^{18}$O bands in our sample and to provide a way to roughly estimate $^{16}$O/$^{18}$O, we have computed  {\sc marcs} model atmosphere spectra for different CO
isotopic abundances \citep{1997A&A...318..521A}. We ran a grid of models
with the stellar parameters $T_{\rm eff}=5000, 5500, 6000$\,K and $\log g = 1.5$ [cgs] and a chemical composition typical of RCB stars
\citep{1997A&A...318..521A}. Values of $^{16}$O/$^{18}$O chosen were, 0, 0.1, 0.33, 1, 3, 10, and infinity (no $^{18}$O). Models with no CO were also run.

The spectra in Figs. 1 and 2 and the {\sc marcs} model spectra are
complex, because, in addition to CO, other species that are abundant in
carbon stars absorb strongly in the observed wavelength interval, as
discussed by \citet{2005ApJ...623L.141C}. In order to estimate the
$^{16}$O/$^{18}$O ratio for each star we first need to measure the
absorption depths of the $^{12}$C$^{16}$O and $^{12}$C$^{18}$O 2-0 and 3-1 bands. To do this, we  started by binning the spectra in Fig.~2 and the {\sc marcs} spectra in intervals of 0.0005~$\mu$m. From the {\sc marcs} spectra with no CO, we then identified wavelength bins in the ``continuum" just shortward and longward of each bandhead where absorption by other species is relatively weak. 
The ratios of the flux densities at the bandheads to the average of those at the ``continuum" wavelengths, corrected for the net absorptions at the bandheads in the {\sc marcs} spectrum with no CO, are shown in the fourth column of Table~2 and provide a first estimate of the isotopic ratio $^{16}$O/$^{18}$O. The values shown are probably overestimates of this ratio where $^{16}$O/$^{18}$O~$<$~1 and underestimates where $^{16}$O/$^{18}$O~$>$~1, because the bands of the more abundant isotopomer are more saturated.

We have interpolated between the {\sc marcs} spectra to estimate the $^{16}$O/$^{18}$O
ratio for each star.  The values are given in the right hand column of
Table~2, and confirm the very low ratios in seven stars. However, the
complexities of the observed and model spectra make determination of the
isotopic ratios problematical, especially in cases such as this, where
individual lines are not resolved. The {\sc marcs} spectra do not match
the observed spectra in detail, adding to the uncertainties. We believe that the derived strengths of the 2-0 bandhead of $^{12}$C$^{18}$O are more reliable than those of the 3-1 bandhead, where the correction for contaminating species was largest, and so we have weighted the ratios of the 2-0 band strengths more highly. Finally, the depths of the CO features in
S~Aps and SV~Sge are greater than those in any of the model spectra that
we generated; for these two stars, we used the 5000~K {\sc marcs} models. For all of
these reasons we estimate the uncertainties in the derived isotopic ratios to be accurate only to a factor of two.

\section{Discussion}

In a prescient paper, \citet{1967MNRAS.137..119W} predicted that HdC stars
would be rich in $^{18}$O. He noted that just before He burning begins
in the core of an intermediate mass star the temperature increases to
values where $^{14}N(\alpha,\gamma)^{18}F(\beta^+\nu)^{18}$O can occur and
subsequently the outer envelope of the star is lost through nova- or
PN-like mass-loss episodes, exposing the $^{18}$O-rich material.  
\citet{2005ApJ...623L.141C}, armed with the detection of a high $^{18}$O
abundance in one HdC star, HD 137613, and using current information about
nuclear reaction rates and stellar interiors, discussed the
nucleosynthesis that could possibly lead to the unusual oxygen and carbon
isotopic abundance ratios in an isolated intermediate mass star. They
noted that the only evolutionary phase that is able to produce the
observed oxygen isotopic ratio as well as satisfy the elemental
abundance constraints for He and CNO, occurs at the onset of He-burning
after completion of H-burning, via the same $\alpha$-capture reaction as
above. For example, in early-AGB stars the $^{18}$O enhancement is found only in a thin layer of the He-burning shell. \citet{2005ApJ...623L.141C}
speculated that HD\,137613 may be a peculiar object that has somehow
managed to strip all of its envelope material precisely down to the narrow
layer where the observed extremely low value of $^{16}$O/$^{18}$O is 
found.

However, the observations reported here demonstrate that an extreme excess of
$^{18}$O is a common and possibly universal feature of HdC and RCB stars.
Even if the above $\alpha$-capture reaction is the source of the enhanced
$^{18}$O, the idea that for a large set of stars post-AGB mass loss
proceeds nearly precisely to the $^{18}$O-enriched layer and then stops is
highly improbable. Another mechanism must be the cause of the enhancement. Moreover, the uniquely low values of $^{16}$O/$^{18}$O in HdC and RCB stars
make it obvious that these two classes of carbon-rich and hydrogen-poor
objects are indeed closely related. Therefore, we suspect that the
isotopic ratios are an important clue in determining the formation pathway
of HdC and RCB stars in general.

Recent studies of V605 Aql and Sakurai's Object imply that a FF origin
for most of the RCB and HdC stars is unlikely \citep{2006ApJ...646L..69C},
based, for example, on the short time that FF stars spend in
the region of the HR diagram where RCB stars are found. Although Sakurai's Object, as a certain representative of the FF scenario, shares a number of
characteristics with the RCB stars, there are some important
differences \citep{1998A&A...332..651A, 1999A&A...343..507A, 2000A&A...353..287A}. Its abundances resemble V854 Cen and the other minority-class RCB stars  \citep{ 1998A&A...332..651A}. However, in Sakurai's Object and probably in V605 Aql,  $^{12}$C/$^{13}$C is low indicating efficient H-burning, while in the RCB and HdC stars it is generally high, thereby excluding the possibility that we are seeing
material processed by efficient H-burning  \citep{2006ApJ...646L..69C}.
Moreover, infrared spectra of Sakurai's Object obtained during 1997-1998 when it had strong CO overtone bands showed no evidence for $^{12}$C$^{18}$O \citep{2002Ap&SS.279...39G}.
Therefore, Sakurai's Object and the other FF stars on the one hand, and most of the RCB and HdC stars on the other hand, are likely to be stars with different
origins. 

As we have discussed before, $^{18}$O can be overproduced in an
environment of partial He-burning in which the temperature and the
duration of nucleosynthesis are such that the
$^{14}N$($\alpha,\gamma$)$^{18}$F($\beta^+$)$^{18}$O reaction chain can produce
$^{18}$O, if the $^{18}$O is not further processed by
$^{18}$O($\alpha,\gamma$)$^{22}$Ne. This has also been suggested by \citet{1986hdsr.proc..127L}\footnote{However, it should be noted that two of the RCB stars, V3795 Sgr and Y Mus, as well as the extreme helium (EHe) stars, show high [Ne/Fe] ratios \citep{2000A&A...353..287A}. 
These stars are too hot to show CO bands so nothing is known about their $^{18}$O abundances.}.
Current understanding of
mixing and nucleosynthesis events in the FF scenario, which leads to
predictions in excellent agreement with Sakurai's Object \citep{2001ApJ...554..298A}, is that
the surface abundance after the FF is dominated
by the intershell material, possibly with the superposed signature of
H-ingestion nucleosynthesis if the FF was a VLTP. In the He-shell flash $^{14}$N is completely
burned into $^{22}$Ne as temperatures are so high that $^{18}$O is destroyed. In
the case of a VLTP, however, there is additional
nucleosynthesis due to the ingestion of H-rich envelope material into
the He-shell flash convection zone. This may separate the upper
H-flash driven convection zone from the original He-shell flash
convection zone. In the upper convection zone, He-burning will 
be terminated and one needs to consider if conditions for $^{18}$O overproduction exist there. However, p-capture nucleosynthesis,
induced by the H-ingestion event, efficiently destroys $^{18}$O and therefore
we do not predict any overabundance of $^{18}$O in the FF case. It should
be mentioned though that the FF scenario involves complicated,
hydrodynamic mixing events that may violate some of the assumptions
made in 1-D stellar evolution. We cannot, therefore, conclusively settle
this question at present.

The above difficulties in producing an extreme $^{18}$O excess together with concerns about the ability of FF events to account for the observed number of RCB and HdC stars have led us to explore an alternative evolutionary origin
for the RCB and HdC stars, namely that of DD mergers. In particular we have considered the onset of He-burning nucleosynthesis in a He-WD/CO-WD merger.
\citet{1984ApJ...277..355W} proposed that an RCB star evolves from the
merger of a He-WD and a CO-WD which has passed through a common
envelope phase. He suggested that as the binary begins to coalesce
because of the loss of angular momentum by gravitational wave
radiation, the (lower mass) He-WD is disrupted. A fraction of the
helium is accreted onto the CO-WD and starts to burn, while the
remainder forms an extended envelope around the CO-WD. This structure,
a star with a helium-burning shell surrounded by a 
$\sim100~R_\odot$ hydrogen-deficient envelope, would resemble that
of an RCB star \citep{1996PASP..108..225C}.
The merging times ($\sim$10$^9$ yr) might not be as long as previously thought, which makes
the DD scenario an appealing alternative to the FF scenario for the formation of RCB stars
\citep{1997ApJ...475..291I, 1996ApJ...456..750I}.  In addition, \citet{2002MNRAS.333..121S} suggested that a
WD-WD merger could also account for the elemental abundances seen in RCB
stars.
\citet{2006ApJ...638..454P} have suggested a similar origin for the EHe stars.

In order to judge the viability of this scenario in the light of our
new observations, and in order to determine the direction of future
more laborious and detailed modeling efforts we have investigated two
separate questions in parallel. We have analyzed the thermodynamic
conditions during the dynamic merger process of the two WDs (Section
\ref{sec:ddmerge}), and we have investigated the necessary conditions
to generate the observed abundance patterns from a
nucleosynthetic point of view (Section \ref{sec:nucl}). As we show in
this section, these two approaches give compatible results, thereby
boosting our confidence that the new observations will eventually be fully
explainable in the WD merger scenario.

\subsection{The Double Degenerate Merger}
\label{sec:ddmerge}

The purpose of this section is to determine the thermodynamic
conditions encountered during the dynamic merger process of a He- and
a CO-WD. We discuss first the conditions that lead to complete,
dynamical tidal disruption of the He-WD, which is the scenario in
which the large overproduction of $^{18}$O may be possible. The critical
parameter is the mass of the He-WD. 
%As we will discuss, very small He-WD masses lead to AM Canum Venaticorum (AM CVn) binaries, in which the CO WD accretes He from the companion at less than a few $10^{-7}\msol$~yr$^{-1}$. Such systems may experience He-shell flashes \citep[see][for recent work on this topic]{2006ApJ...640..466B} during which $^{18}$O is destroyed. 
After discussing the
conditions for tidal disruption, we examine the thermodynamic
conditions during disk accretion to determine if they allow the types of nucleosynthesis that might result in the observed overproduction of
$^{18}$O.

\subsubsection{Tidal Disruption}

To determine the extent and structure of the disk formed in the
merger, we must examine the merger evolution.  The merger process
begins when the He-WD is brought so close to its companion (either by
magnetic braking or gravitational wave emission) that it overfills its
Roche radius.  The orbital separation at which a WD of this radius
will overfill its Roche radius is given by
\citet{1983ApJ...268..368E}: 
\begin{equation} 
A_0 = R_{\rm He} \frac{0.6 q^{2/3} + ln(1+q^{1/3})}{0.49q^{2/3}}, 
\end{equation} 
where $q=M_{\rm He}/M_{\rm CO}$ with $M_{\rm CO}$ equal to the mass of the
CO-WD and $M_{\rm He}, R_{\rm He}$ are set to the mass and radius of the
He-WD, respectively.  Assuming a $\Gamma=5/3$ polytrope approximation
of a WD equation of state, we derive an equation for the radius of the
He-WD \citep{1972ApJ...175..417N}:
\begin{equation} R_{\rm He} = 10^4 (M_{\rm He}/0.7 M_\odot)^{-1/3}
[1-M_{\rm He}/M_{\rm Ch}]^{1/2} (\mu_e/2)^{-5/3} {\rm km,}
\label{eq:radhe} \end{equation} where $M_{\rm Ch} \approx 1.4$ is the
Chandrasekhar mass and $\mu_e$ is the mean molecular weight per
electron of the WD.

As matter flows from the He-WD onto a disk around its companion, both
the Roche-overflow radius of the WD and the orbital
separation of the binary change.  The degenerate nature of the He-WD causes it to expand as it loses mass.  If orbital angular
momentum were conserved, the orbital separation would also increase as
the low-mass He-WD accretes onto its heavier companion.  The accretion
process could then follow one of two possible paths: the He-WD radius
increases faster than the orbital separation and the WD is disrupted over the
course of one or two orbital periods, or the orbital separation increases
faster than the WD radius and the He-WD accretes onto its
companion on a timescale set by magnetic braking.  To determine
the path our objects follow, we compare the expansion of both the
He-WD and the binary separation.

However, orbital angular momentum is not conserved; a fraction of it is transferred into angular momentum of the disk.
Assuming no mass is ejected, the angular momentum as a function of the
evolving WD mass is 
\citep{1999ApJ...520..650F,1992ApJ...391..246P}:  \begin{equation}
\frac{A}{A_0} = \left ( \frac{M_{\rm He}}{M^0_{\rm He}} \right )^{C_1}
\left ( \frac{M_{\rm CO}}{M^0_{\rm CO}} \right )^{C_2}, \end{equation}
where $C_1 \equiv -2 + 2 j_{\rm disk}$ and $C_2 \equiv -2 - 2 j_{\rm
disk}$ and $j_{\rm disk}$ is the specific angular momentum of the
accretion disk material \citep[see][for details]{1999ApJ...520..650F}.
\citet{1999ApJ...520..650F} found for the case of WD-black hole
mergers that more than half of the angular momentum in the accreting
material found its way into the disk or into the black hole spin (that
is, $j_{\rm disk} \gtrsim 0.5$).  It is likely the larger disks
produced by these low-mass WD mergers will carry even more angular
momentum.

The orbital separation as a function of He-star mass for four separate
binary systems is shown in Fig.~\ref{fig:merger}.  The solid
curves show the path we expect the binaries to follow as the He-star
loses mass (going right to left in decreasing He-star mass).  The four
curves denote four different values of $j_{\rm disk}$.  The dotted
curve shows the separation for Roche-lobe overflow for this binary
(again, a function of the He-star mass).  There are two things to note
in this figure. First, if $j_{\rm disk}>0.5$, the
separation of the binary system will always be less than the
separation for Roche-lobe overflow until the star is disrupted ($\sim
0.1\,M_\odot$). This means that when the WD starts accreting, it will
not stop until the companion WD is completely disrupted.  The
accretion process should proceed on a few orbit timescale: $\sim 1000$\,s.
Second, the orbital separation does not increase significantly during
this accretion process for such high values of $j_{\rm disk}$.  Tidal
effects will send some material as far out as twice the initial
separation (but very little will be ejected - see
\citet{1999ApJ...520..650F}).  But the bulk of the accreting material
will not span more than twice the initial orbital separation at the
onset of Roche-lobe overflow.

\subsubsection{Disk Accretion}

Based on WD merger calculations by \citet{1999ApJ...520..650F}, we expect
most of the matter in the He-WD to accrete onto its CO-WD companion.  The
He-WD is disrupted nearly instantaneously, forming a disk with a total radius
of nearly 2 times the separation at the onset of Roche-lobe overflow.  
The accretion rate onto the CO-WD will definitely vary with time, but to obtain a preliminary estimate of it and of the temperature, we calculate the average value for these quantities.

Under the $\alpha$ disk prescription for transporting angular momentum in
the disk formed by the WD disruption, we can estimate the accretion time:
$t_{\rm acc} = P_{\rm disk}/\alpha$ where $P_{\rm disk}$ is the orbital
period of the disk and $\alpha$ is the viscosity of the disk.  
Fig.~\ref{fig:accretion} shows this accretion time as a function of the
mass of the CO-WD assuming $\alpha=10^{-3}$.  The corresponding mass
accretion rate ($\dot{M}_{\rm acc}$), roughly $M_{\rm He}/t_{\rm acc}$, is
also plotted in Fig.~\ref{fig:accretion}.  The entire disk is likely to
accrete in a few days, with accretion rates as high as
150\,M$_\odot$\,yr$^{-1}$.

If this material accreted in a steady system, it would build up an
atmosphere on the WD.  Accretion atmospheres have been studied in detail
for the case of neutron stars \citep{1989ApJ...346..847C,1996ApJ...460..801F,1993PhR...227..157C}. \citet{1996ApJ...460..801F} derived the entropy, $S$, of this atmosphere as a function of the compact remnant mass, in this case, the WD mass ($M_{WD}$),
position of shock ($r_{\rm shock}$)\footnote{This is initially the CO-WD
radius but it may move outward an order of magnitude}, and the accretion
rate ($\dot{M}_{\rm acc}$).  The pressure of this accreting material will be
nearly dominated by photons.  For a $\gamma=4/3$, radiation dominated
gas, the entropy is given by: \begin{equation} S=27.9 M_{WD}^{7/8}
\dot{M}_{\rm acc}^{-1/4} r_{\rm shock}^{-3/8} \end{equation} where the
$M_{WD}$ is in $M_{\sun}$, $\dot{M}_{\rm acc}$ is in $M_{\sun}$ yr$^{-1}$ and
$r_{\rm shock}$ is in $10^{10}$\,cm.  As the atmosphere builds, the
entropy at higher layers will decrease, and this negative entropy gradient
will lead to rapid convection, producing a flatter entropy profile on
timescales inversely proportional to the Brunt-V\"ais\"ala frequency,
roughly 1-100\,s. But the extent of this atmosphere, and hence, the lower
limit on the entropy, is difficult to determine without actual
calculations (indeed, given the dynamic nature of this accretion, such a
simple picture may not even be correct).  Typical initial entropies are
roughly 10-30 erg K$^{-1}$ for this accreting atmosphere, but it is possible that the
convection can drive the value down by a factor of $\sim2$.  We can derive
a temperature and a density of the atmosphere as a function of entropy:
\begin{equation} T=2.262 \times 10^8 S^{-1} (R/10^{10} \,{\rm cm})^{-1}
{\rm K}, \end{equation} and \begin{equation} \rho= 390 (S/10 {\rm
k_B~nucleon^{-1}})^{-4} (R_{\rm CO}/10^9 {\rm cm})^{-3} {\rm g~cm^{-3}}.
\end{equation} The corresponding temperature for the atmosphere at the
surface of the CO-WD lies in the $1-2\times 10^8$\,K range if we assume
the highest value for the entropy.  The temperature could increase if
mixing brings lower entropy material down, but is likely not to exceed 4$\times
10^8$~K. The high temperatures and short dynamic nature of the merger process are supported  by multi-dimensional SPH simulations as well \citep{2004A&A...413..257G}.

 \subsection{Implications from Nucleosynthesis}
\label{sec:nucl}

Now that we have a rough physical picture of the merger process and
the corresponding thermodynamic conditions we turn to the
nucleosynthesis part of the problem. We ask what conditions are
required for nucleosynthesis to produce the observed abundances.

\subsubsection{Observational Signatures of HdC and RCB Stars}

Typical RCB abundances are characterized by extreme hydrogen deficiency,
enrichment relative to Fe, of N, Al, Na, Si, S, Ni, the s-process elements
and sometimes O \citep{1986hdsr.proc..127L,2000A&A...353..287A}. 
Based on their abundance characteristics,
RCB stars can be divided into a homogeneous majority and a
diverse minority class, the latter distinguished by,
among other things, lower metallicity, generally less striking H-deficiency, 
and extreme abundance ratios such as Si/Fe and S/Fe \citep{2000A&A...353..287A}. The isotopic carbon ratio, $^{12}$C/$^{13}$C, where measured, is very large ($>500$) in most RCB stars. 
One exception is the minority-class RCB, V CrA, where $^{12}$C/$^{13}$C is estimated to be between 4 and 10 \citep{2005ASPC..336..185R}.
The elemental abundances of the HdC stars are similar to those of the majority of RCB stars \citep{1967MNRAS.137..119W,1997A&A...318..521A,2002BaltA..11..249K}.  
Lithium is seen in four of the majority-class RCB stars and at least one HdC star, HD 148839 \citep{1996ASPC...96...43R}. Also, although difficult to estimate, C/He is believed to be high in both classes of stars \citep[$\sim$1-10\%,][]{2000A&A...353..287A}.

The surface compositions of HdC and RCB stars are extremely He-rich (mass
fraction 0.98), indicating that the surface material has been processed
through H-burning. After H-burning via the CNO cycle, N is by far the
most abundant of the CNO elements, because \nvi\ has the smallest nuclear
p-capture cross-section of any stable CNO isotope involved. However, the majority of RCB stars has 
$\log{C/N} = 0.3$\footnote{There is still an unresolved C-problem in the spectral abundance analysis. The value, $\log{C/N} = 0.3$, for the majority RCB stars is only true if the derived C abundance rather than the input
C abundance is adopted, which rather would give log C/N~=~0.9 \citep{2000A&A...353..287A}.} 
and $\log{N/O} = 0.4$ by number, equivalent to mass ratios of $C/N =
1.7$ and $N/O = 2.2$ \citep{2000A&A...353..287A}. The N/O ratio represents the average for the majority RCB stars although the individual stars show a large scatter. Thus C is the most
abundant and O the least abundant CNO element. These abundances are
consistent if the material at the surfaces of
HdC and RCB stars experienced a small amount of He-burning, as for
example at the onset of a He-burning event that is quickly
terminated. This partial He-burning would not significantly deplete He,
but could be sufficient for some of the \nvi\ to be processed into \oac.
%Other abundance signatures of RCB and HdC stars are a high \czw/\cdr\
%ratio ($>500$) and s-process enhancement \citep{asplund:00}. Li
%overabundances have also be observed in some objects of this class.  
At the onset of He-burning, \cdr\ is the first $\alpha$-capture
reaction to be activated because of the large cross-section of $\cdr(\alpha,\n)\ose$.  Thus, a large $\czw/\cdr$ ratio and
enhanced s-process elements are both consistent with partial He-burning.

\subsubsection{Parametric Nucleosynthesis Model}

We constructed a parametric nucleosynthesis model for a quantitative estimate of the partial He-burning nucleosynthesis scenario 
(\abb{fig:t-H-He-T}). We consider a
single zone nucleosynthesis calculation with a solar initial abundance
distribution. The density is assumed to be $2 \times 10^3~\mem{g~cm^{-3}}$
and the temperature initially is $5 \times 10^7$\kelv. We follow the
nucleosynthesis with a nuclear network code until the H has been
burned to a mass fraction $<10^{-4}$. This situation is reached
after approximately $125\jahre$. This first phase in our model
represents H-burning, e.g., in the core or in a shell, that produces the
He in the He-WD. The assumed density is unrealistically high,
because, for simplicity we do not adjust the density in different
phases of the model. In practice, varying the density will only affect
the timescale associated with each burning phase. Here, the first
phase serves only one purpose: to easily arrive at an abundance
distribution that reflects that in a He-WD.

At the end of the H-burning phase, we increase the temperature to
switch smoothly into the He-burning regime (\abb{fig:t-H-He-T}). In
our scenario, this temperature rise will occur as a result of the
accretion of the He-WD onto the more massive CO-WD. Here we assume a
temperature rise of $1\%$~yr$^{-1}$. The timescale is not important here
as everything scales with the density.  The purpose is only to see how
the relative abundances change as a function of temperature.

We stop the increase when T~=~$1.65~\times~10^8$\kelv, which occurs at
$\approx245~\jahre$. For even slightly larger maximum temperatures,
He-burning quickly reduces the abundances of He and $^{14}$N to much smaller
values than observed in the HdC and RCB stars. Lower temperatures require a
much longer time for \nvi\ to be transformed into \oac.  From the merger considerations in the previous section, we
anticipate that the period of He-burning cannot be very long if it is
a result of the dynamical merging process. The exact timescale
depends on the density evolution. After stopping the temperature increase, we
allowed the network to continue for another $255\jahre$ until abundances
are reached that clearly disagree with those seen in the HdC and RCB
stars.

\subsubsection{Results}

\abb{fig:t-H-He-T} (upper right panel) shows the evolution of the
dominant CNO isotopes.  Initially the CNO H-burning equilibrium
ratios, including the large \nvi\ abundance are established. Then, as
the temperature increases, a modest increase in the \ose\ abundance
can be seen (at 220 yr), resulting from the activation of
$\cdr(\alpha,\n)\ose$. Thereafter \ose\ remains largely unchanged
because the temperature does not become high enough for
$\czw(\alpha,\gamma)\ose$. However, just after \cdr\ burning, \nvi\
begins undergoing $\alpha$-capture, producing $\oac$. This reaction
very quickly makes \oac\ the most abundant of the CNO isotopes, a
situation that lasts until $\oac$ begins capturing $\alpha$ particles
to produce \nezw\, at which time it is overtaken by \czw\ whose
abundance has been increasing since $t=~230~\jahre$ because of the
triple-$\alpha$ reaction.

The evolution of the elemental C/N and N/O ratios (lower left panel)
is qualitatively consistent with nucleosynthesis after full completion
of H-burning and at the beginning of He-burning.  The N/O ratio
decreases when the temperature is high enough to start He-burning, and
eventually reaches the mean observed value at $\sim250\jahre$. The C/N
ratio increases, but reaches the observed mean value somewhat
later. However, both times are consistent with a very high and only
slightly decreased He abundance at this early time of He-burning.

The most pertinent information in Fig.~5 is the evolution of the \ose/\oac\ and \czw/\cdr\ ratios (lower right panel). The enormous production of
\oac\ from \nvi\ leads to a rapid decrease of the \ose/\oac\ ratio at
the onset of He-burning, which passes through the observed value for
HD~137613 at around $250\jahre$. The \czw/\cdr\ ratio increases to the
point that it exceeds its observed lower limit of 500
\citep{1977PASJ...29..711F} at a slightly earlier time.

\subsubsection{Admixture of H} 

It is intriguing that most of the observed abundances of the HdC and RCB
stars that have been discussed so far seem consistent with a
nucleosynthetic origin at the onset of He-burning. One must keep in mind that this simulation consists of only a single zone, and that the temperature evolution is roughly assumed to demonstrate the various stages of nucleosynthesis in a DD merger. Nevertheless, one may complain that the two
observations are not well matched. The C/N ratio increases too late, when
\ose/\oac\ and N/O
%(and maybe the \czw/\cdr\ ratio \texttt{a lower limit is a lower limit, 
%isn't it? But maybe the observations can exclude $10^{10}$? {\it A lower 
%$limit means, nothing is detected so it can be zero.}}) 
have already reached values far from those observed. The other
inconsistency is the observed enhancement of s-process elements. In
our model, the only source of neutrons is the small amount of \cdr\
from H-burning via the CNO cycle. This is insufficient by far to
produce significant s-process enhancements. In addition, a good
match for  the CNO elemental and isotopic abundances is only obtained for a short period. Possibly the observed
values are the result of a characteristic termination timescale for
this process and the associated nucleosynthesis.
The admixture of H could lead to an enrichment of $^{14}$N \citep{1996ASPC...96...43R, 2000A&A...353..287A}.

However, the CO-WD retains a small H-rich envelope,
$10^{-3}$-$10^{-4}M_{\sun}$ depending on the WD mass. The accreting material
will be shocked and heated, quickly leading to violent He-burning, which
will induce convection. In addition, the accreted layer will be perturbed
by the ongoing dynamic merging phase. All this will help to bring the
original H-rich WD envelope up into the accreted He-burning material. In
particular, the fact that the accreted elements have larger atomic masses
make mixing of the H-rich CO-WD envelope into the accreted He-rich
material inevitable.

The question is when and how much H-rich material will be added. With our one-zone model we can only crudely explore the effect of such an
H-admixture. In reality, there will be
multi-zone mixing and burning at a range of densities and
temperatures. Nevertheless, if the temperatures of the hottest layers
are about right, one can obtain a zeroth-order approximation to what
really happens. The admixture of a substantial fraction of the H-rich
CO-WD envelope has to occur at a time when some \czw\ has already been
produced through He-burning, because a significant new source of \czw\
is required for additional p-captures to produce the \cdr\ that can
serve as a neutron source for the s-process. From \abb{fig:t-H-He-T}, it is
clear that triple-$\alpha$ reactions are only activated at a
temperature of $1.6 \times 10^8$\kelv\ at the assumed density. Then, the
admixture must occur after that temperature has been reached.

To test these ideas, we selected a CO-WD envelope mixture to be blended into the He-burning accreted material. The mixture comes from an actual post-AGB stellar evolution calculation with a core mass of 0.6 $M_{\sun}$ and an initial ZAMS mass of 2 $M_{\sun}$. It has enhanced \czw\ ($\mem{X_C}=1.4 \times 10^{-2}$) and
\ose\ ($\mem{X_O}=1.5 \times 10^{-2}$), considerably larger than our initial
choice of solar abundances ($\mem{X_C}=3 \times 10^{-3}$,
\mem{X_O}=$8.3 \times 10^{-3}$) due to several third dredge-up events during
the TP-AGB evolution phase preceding the WD formation. 
The accreted He-WD material has a mass of order 0.1 $M_{\sun}$, while the CO-WD
envelope has a mass of the order $5 \times 10^{-4}$ $M_{\sun}$. Thus, we admix the
envelope in a ratio 1:0.005 and renormalize to unity. We examine the effect of
this admixture at the time when the \czw\ abundance reaches a mass
fraction of $5 \times 10^{-3}$.

\abb{fig:nucl-Hadm} shows the results of
the network calculation. As the H-rich material is admixed at this early
stage of He-burning, it encounters a large \oac\ abundance. Through two
successive p-captures, $\oac(\p,\alpha)\nfu(\p,\alpha)\czw$, most of the
\oac\ is transformed into \czw. \fne\ is formed by $\oac(\p,\gamma)\fne$
and, because of the large abundance of \hevi\, also through
$\nfu(\alpha,\gamma)\fne$. 
%\texttt{I would have to look at the exact
%fluxes, this is admittedly some educated guesswork, and probably correct
%in the end.}
 Temporarily, the \fne\ abundance exceeds the solar value by
two orders of magnitude.  We note that
lines of F~I were recently detected for the first time in visible spectra of a group of extreme He stars \citep{2006ApJ...648L.143P}. 
The \czw\ nuclei, both
freshly produced from triple-$\alpha$ (primary), and coming from \oac\
make \cdr\, in turn producing neutrons and \ose~through $\alpha$-capture:
$\czw(\p,\gamma)\ndr(\beta^+)\cdr(\alpha,\n)\ose$.

Due to the burning of \cdr, the \ose\ abundance is
increasing quickly.  The abundance of \nvi\ also increases because of
$\cdr(\p,\gamma)\nvi$.
%There are several other reactions that may play a role here (and need
%more investigation), e.g., the small decrease of \nvi\ after the initial
%increase, when \ose\ is still increasing a little bit, is probably
%associated with the $\nvi(\n,\p)\cvi$ reaction. \cvi\ has a half-life of
%5000\jahre\ and will feed back into the \oac\ abundance through $\alpha$
%capture or \nfu\ through a \p-capture. 
The admixture takes place just after the C/N ratio has increased
significantly above the observed mean value. Due to the burning of
\czw\ and the creation of \nvi, the C/N ratio drops below the
observed value again. The opposite happens to the N/O ratio. At the
time of the admixture, \oac\ dominates the O elemental abundance. Its
quick decrease plus the increase in \nvi\ lead to a rapid increase in
the N/O ratio above the observed ratio.

The effects of the admixture on the \czw/\cdr\ and \ose/\oac\ ratios
are dramatic. Due to the burning of \oac\ and the increase in \ose\
from the $\cdr(\alpha,\n)\ose$ reaction, the \ose/\oac\ ratio quickly
rises beyond the observed values. The \czw/\cdr\ ratio decreases,
partly due to the admixture of \cdr\ from the CO-WD envelope (the first
jump down to $\log \czw/\cdr \approx 4$) and then partly due to the
production of \cdr\ from $\czw(\p,\gamma)$.

The effect on s-process elemental abundances should also be profound. Due to 
the production of the \cdr\ at high He-burning 
temperatures, neutrons are rapidly
released. In this particular admixture test, we find $\tau_\mem{exp} =
0.13~\mem{mbarn}^{-1}$ and a peak neutron density of $N_\mem{n} >
10^{11}~\mem{cm}^{-3}$. Thus, the neutron exposure is
in the range that could explain the observed s-process elements \citep{2000A&A...353..287A}. However, the neutron
density is far too high to produce an s-process signature.

In summary, it seems that the admixture of H can aid in solving the delay in the match
of the elemental and isotopic ratios. It also has, in general, the
effect of reversing the effect of too much He-burning. For example, the
He-burning leads to the decrease of \ose/\oac\ to values
far below the lowest observed. H-admixture has the effect of reversing
this trend and increasing the O isotopic ratio again. The same can be
observed for the other abundance ratios. We therefore speculate that
H-admixture in a realistic, highly mixed accretion environment can
help to stabilize the nucleosynthesis results leading to the
significant coherence of abundance patterns observed in HdC and RCB
stars.

\section{Conclusions}

Observations of the first overtone CO bands in cool HdC and RCB stars have resulted in two significant findings: (1) in all HdC stars and most or all RCB stars where oxygen isotopic ratos can be measured, $^{18}$O is enhanced relative to $^{16}$O by factors of 100--1000 compared to standard Galactic values; and (2) the lowest $^{16}$O/$^{18}$O ratios are found in the HdC stars. 
Our models show that nucleosynthesis in the dynamically accreting material of CO-WD--He-WD mergers, previously suggested as the source of RCB stars, may provide a suitable environment for significant production of $^{18}$O and the very low $^{16}$O/$^{18}$O values as observed. The suggestion that FF events produce RCB stars appears to be weakened due to a mismatch of oxygen (this paper) and carbon isotopic abundances, as well as by the apparent inconsistency in timescale of the RCB-mimicking portion of the FF episode and that required to account for the number of RCB stars \citep{2006ApJ...646L..69C}. Therefore, we have explored the idea that HdC and RCB stars originate in the mergers of CO- and He-WDs. 

In the WD merger, it is partial He-burning at just the right temperature and duration that is the likely generator of small \ose/\oac\ ratios. In addition, the high values of $\czw/\cdr$ observed in HdC and RCB stars can probably be reproduced at the same time. The observed s-process abundances in these stars as well as the delayed increase of the C/N ratio to the observed mean value motivated the investigation of the
effects of admixing H-rich material. This material would, in the DD
merger/accretion scenario, be the thin CO-WD envelope.  We found that
qualitatively the instantaneous admixture of H-rich material resets
all observables back toward the observed values. The fact that all of
the H-rich material is added instantaneously leads to artefacts such
as the extremely large neutron density, as well as the very small
\czw/\cdr\ ratio. In reality, the admixture will take place
over a finite time and in smaller quantities. A slower rate of admixing could produce the observed abundances.

It is interesting to note that some RCB stars show enhanced Li
abundances, as does the FF star Sakurai's Object \citep{1986hdsr.proc..127L,1998A&A...332..651A}. As shown by \citet{2001NuPhA.688..221H}, Li enhancements are consistent with the FF
scenario. However, the production of $^{18}$O requires temperatures
large enough to at least marginally activate the
$^{14}$N$(\alpha,\gamma)$ reaction. The $\alpha$ capture on Li is eight
orders of magnitude more effective than on $^{14}$N. For that reason the
simultaneous enrichment of Li and $^{18}$O is not expected in the WD merger scenario. This is an
important argument against the FF evolution scenario as a progenitor
of RCB and HdC stars with excess of $^{18}$O. 
The presence of Li in 
some RCB stars, as well as in Sakurai's Object,
is extremely difficult to explain with a WD merger. Since
$^{18}$O strongly supports the DD merger/accretion scenario for most of the stars studied so far, 
the obvious solution is that there are (at least) two 
evolutionary channels leading to RCB, HdC and EHe stars, perhaps
with the DD being the dominant mechanism. Unfortunately the
division between majority- and minority-class RCB stars does not lend itself naturally to this explanation,
since Li  has only been detected in
the majority group \citep{2000A&A...353..287A}.

Contrary to the work of \citet{2002MNRAS.333..121S}, who assumed an
accretion rate of 10$^{-5}\msun/\jahre$ to model WD mergers with a
hydrostatic evolution code, our estimate of the merger process
leads to a picture of a rapid, dynamic event that only lasts a few
days, and during which high temperatures can be obtained. In the merger scenario, rapidity appears to be essential to producing the conditions required to account for our observations of extremely abundant $^{18}$O.

Understanding the RCB and HdC stars is a key test for any theory that aims
to explain hydrogen deficiency in post-AGB stars.
Confirmation of the WD merger scenario
will allow the use of RCB and HdC stars as probes for WD merger
simulations. The ability to model the rates of these low-mass WD mergers
will help us to understand the rates of more massive mergers that make
type Ia supernovae \citep{2005ApJ...629..915B}.

%\section{Summary}

\acknowledgments

Most of these observations were obtained during program GS-2005B-Q-20 at the Gemini Observatory, which is operated by AURA under a cooperative agreement
with the NSF on behalf of the Gemini partnership of Argentina,
Australia, Brazil, Canada, Chile, the United Kingdom and the United
States of America. We are indebted to the Gemini staff for its expertise in obtaining these data. We also thank the staff of UKIRT, which is operated by the Joint Astronomy Centre on behalf of the U.K. Particle Physics and Astronomy Research Council. TRG's research is supported by the Gemini Observatory, which is operated by the Association of Universities for Research in Astronomy,
Inc., on behalf of the international Gemini Partnership. This work was carried out in part under the auspices of the National Nuclear Security Administration of the
U.S. Department of Energy at Los Alamos National Laboratory under
Contract No. DE-AC52-06NA25396, and funded by the LDRD program
(20060357ER). This work is LANL report LA-UR-07-0198.
This work was carried out in part in collaboration with the NSF funded Joint Institute for Nuclear Astrophysics (JINA).

%\clearpage

\bibliography{/Users/gclayton/projects/latexstuff/everything2,/Users/gclayton/projects/rcb/emmy/gemini/DDMERGO18_060926/astro}

\clearpage

\begin{deluxetable}{llcrrrr}
\tablecolumns{7}
\tablewidth{0pc}
\tablecaption{CO Bandhead Absorption Depths in RCB and HdC
Stars$^{a}$}
\tablenum{1}
\tablehead{\colhead{Star}&\colhead{Tel.}&\colhead{Exp.}&\multicolumn{2}{c}{$^{12}$C$^{16}$O} &\multicolumn{2}{c}{$^{12}$C$^{18}$O}\\
\colhead{}&\colhead{}&\colhead{Time (s)}&\colhead{2-0}&\colhead{3-1}&\colhead{2-0}&\colhead{3-1}}
\startdata
HD 175893 & GS    & 480 &    0.06 &    0.07 &       0.15 &       0.11\\
HD 182040 & GS    &  56 &    0.03 &    0.05 &       0.05 &       0.06\\
HD 137613 & UKIRT & 480 &    0.12 &    0.20 &       0.18 &       0.21\\
WX CrA    & GS    & 480 &    0.18 &    0.20 &       0.15 &       0.13\\
S Aps     & GS    & 160 &    0.54 &    0.51 &       0.26 &       0.15\\
SV Sge    & GS    & 80  &    0.51 &    0.47 &       0.24 &       0.11\\
ES Aql    & GS    & 480 &    0.35 &    0.31 &       0.09 &       0.07\\
Z Umi$^{b}$& Bok  & 60  &    0.34 &    0.34 &       0.07 &       0.02\\
U Aqr     & GS    & 720 &    0.16 &    0.24 &       0.03 &    $<$0.02\\
HD 148839 & GS    & 480 & $<$0.05 & $<$0.03 &    $<$0.03 &    $<$0.02\\
HD 173409 & GS    & 960 & $<$0.03 & $<$0.02 &    $<$0.02 &    $<$0.02\\
\enddata
%\tablenotetext{a}{HD 137613 data from UKIRT; all other stars except Z UMi observed at Gemini South.}
\tablenotetext{a}{Based on flux ratios at rest vacuum wavelengths 2.2930
and 2.2940~$\mu$m, 2.3215 and 2.3235~$\mu$m, 2.3480 and 2.3495~$\mu$m, and
2.3770 and 2.3790~$\mu$m for $^{12}$C$^{16}$O 2-0 and 3-1 bands and
$^{12}$C$^{18}$O 2-0 and 3-1 bands, respectively.}
\tablenotetext{b}{Spectrum originally published by
\citet{2005AJ....130..256T}.}
\end{deluxetable}

\begin{deluxetable}{lllcc}
\tablecolumns{5}
\tablewidth{0pc}
\tablecaption{Estimated Isotopic Oxygen Abundances}
\tablenum{2}
\tablehead{\colhead{Star} & \colhead{HdC/RCB} & \colhead{T$_{eff}$}
& \colhead{abs(C$^{16}$O)/abs(C$^{18}$O)} &
\colhead{$^{16}$O/$^{18}$O$^{a}$}}
\startdata
HD 175893    & HdC &       5500 & 0.4 & 0.2\\
HD 182040    & HdC &       5600 & 0.6 & 0.3\\
HD137613&HdC&5400&0.7&0.5\\
WX CrA       & RCB &       5300 & 1.3 & 1\\
S Aps        & RCB &       5400 & 2.5 & 4\\
SV Sge       & RCB &       4000 & 2.5 & 4\\
ES Aql       & RCB & 5000 & 3 & 6\\
Z UMi&RCB&5000&$\geq$4&$\geq$8\\
U Aqr        & RCB &       6000 & $\geq$6 & $\geq$12\\
HD 148839    & HdC &       6500 & \nodata & \nodata\\
HD 173409    & HdC &       6100 & \nodata & \nodata\\
\enddata
%\tablenotetext{a}{from \citet{1997A&A...318..521A}, \citet{1990MNRAS.247...91L},\citet{2001A&A...369..178B}.}
\tablenotetext{a}{Estimated uncertainty is a factor of two.}
\tablerefs{\citet{1997A&A...318..521A,1990MNRAS.247...91L,2001A&A...369..178B}}
\end{deluxetable}

\clearpage

\begin{figure}
\figurenum{1} 
\begin{center}
\includegraphics[width=4in,angle=0]{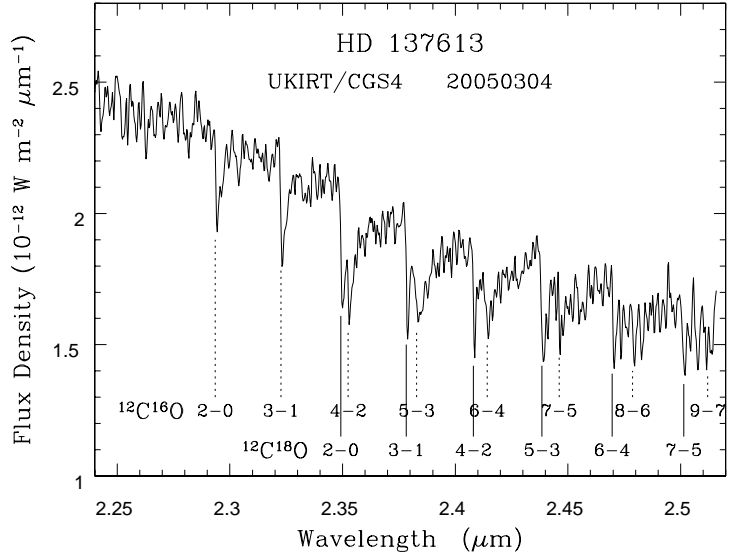}
%\includegraphics*[width=3.25in]{SNextracted/doug2.eps}
%\epsscale{1.0}
%\plottwo{f1a.eps}{f1b.eps}\\
%\epsscale{2.0}
%\plottwo{f1c.eps}{f1d.eps}\\
%\epsscale{0.5}   
%\plotone{f1e.eps}
\end{center}
\caption{UKIRT spectrum of HD 137613. The wavelengths of the bandheads of 
$^{12}$C$^{16}$O and $^{12}$C$^{18}$O are indicated.}
\end{figure}

%\clearpage

\begin{figure}
\figurenum{2}
\begin{center}
\includegraphics[width=5in,angle=0]{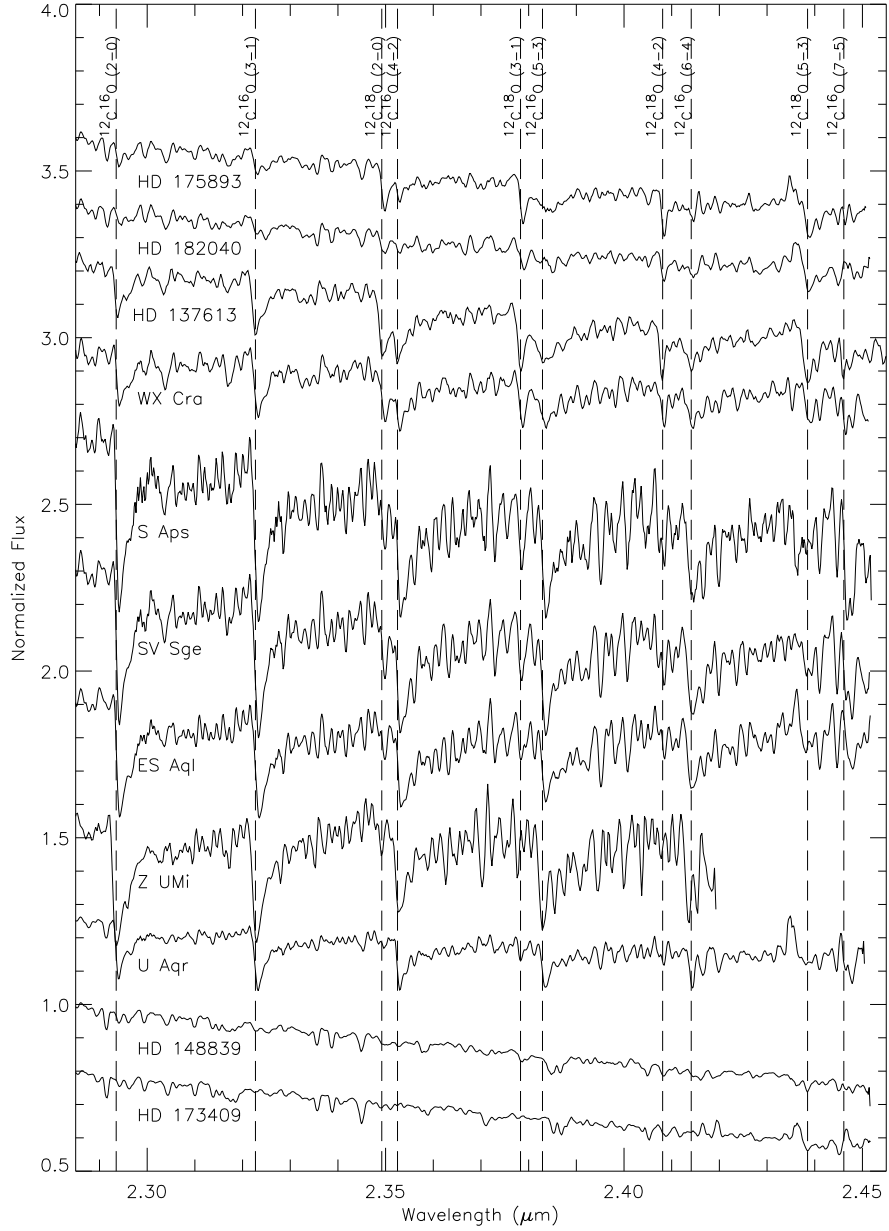}
%\includegraphics*[width=3.25in]{SNextracted/doug2.eps}
%\epsscale{1.0}
%\plottwo{f1a.eps}{f1b.eps}\\
%\epsscale{2.0}
%\plottwo{f1c.eps}{f1d.eps}\\
%\epsscale{0.5}
%\plotone{f1e.eps}
\end{center}
\caption{2.28--2.45$ \mu$m spectra of RCB and HdC stars, with wavelengths of $^{12}$C$^{16}$O and 
$^{12}$C$^{18}$O bandheads indicated by vertical lines.   
The stars are ordered by the strengths of the $^{12}$C$^{18}$O bands. The first three spectra at the top and the last two at the bottom are of HdC stars. The rest are of RCB stars.
 }
\end{figure}

\begin{figure}
\figurenum{3}
\begin{center}
\includegraphics[width=3.5in,angle=0]{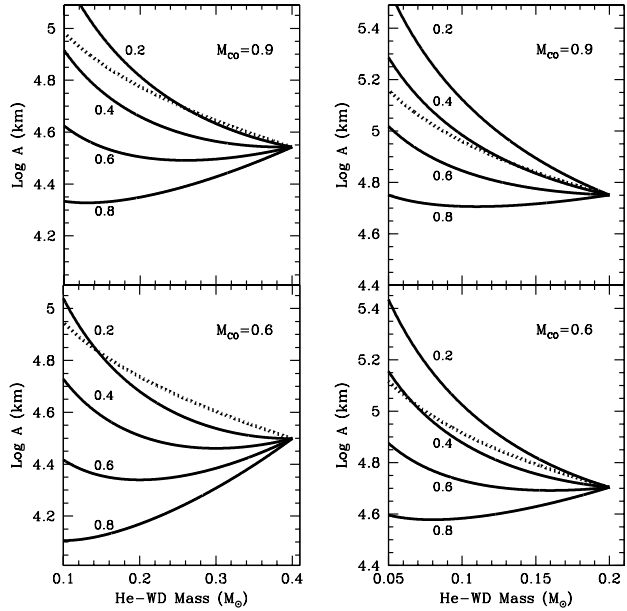}
\end{center}
\caption{Evolution of  a WD binary as a function of the masses of
the two stellar components.  The start of the evolution is where all
the lines meet at a point (right side of the plots).  The solid lines
show the evolution of the orbital separation of the binary as mass
transfers from the He-WD onto the CO-WD.  The different curves correspond to varying amounts of angular
momentum in the disk as a function of the mean binary angular
momentum: 0.2 means very little angular momentum goes into the disk,
most stays in the binary whereas 0.8 means that the accreting material
retains most of its angular momentum as it forms an accretion disk.
Models of WD/BH mergers \citep{1999ApJ...520..650F} and NS/BH mergers \citep{1999ApJ...527L..39J}
all suggest that this number is at least $\sim 0.5$.  The dotted
line denotes the binary separation below which Roche-lobe overflow
occurs.  Because degeneracy dominates the pressure of the He-WD, as it loses mass, it expands, and the separation at which
Roche-lobe overflow occurs also increases.  We start the evolution
with the orbital separation set to the Roche overflow separation (the
onset of Roche-lobe overflow).  If the subsequent evolution causes the
separation to drop below the Roche overflow separation, the accretion
will develop dynamically and will lead to the disruption of the
He-WD on an orbital timescale.  The four panels correspond to two
CO-WD masses (top - 0.9 M$_\odot$, bottom - 0.6 M$_\odot$) and two He-WD masses (left - 0.4 M$_\odot$, right - 0.2 M$_\odot$).
\label{fig:merger}}
\end{figure}
\clearpage

\begin{figure}
\figurenum{4}
\begin{center}
\includegraphics[width=3.5in,angle=0]{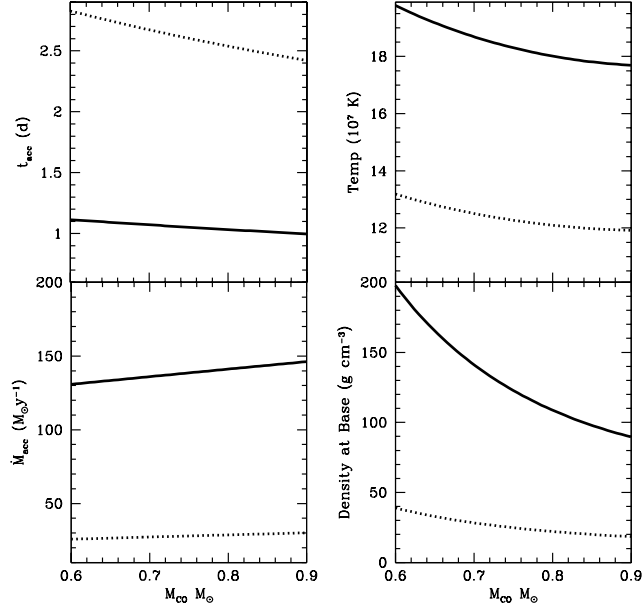}
\end{center}
\caption{Properties of the accretion atmosphere as a function of the
C/O mass.  The solid and dotted curves correspond to 0.4 and 0.2\,M$_\odot$
He stars, respectively.  The top left panel shows the accretion
timescale, the bottom left panel shows the corresponding accretion
rate.  The largest uncertainty for both of these terms is the choice for
the viscosity (we have assumed $\alpha \approx 10^{-3}$).  This value
could be a factor of 100 larger, producing accretion timescales a
factor of 100 shorter and accretion rates a factor of 100 larger.  If
we assume this accretion builds an equilibrium atmosphere on top of
the CO-WD, we can then estimate the temperature and density
at the base of this atmosphere.  This calculation depends upon the
size of the atmosphere and also the accretion rate.  The latter is the
biggest uncertainty and could lead to an increase in the temperature
of, at most, a factor of 2, and an increase in the density of roughly a
factor of 10.  We note, however, that nucleosynthetic constraints
argue for our standard calculation.
\label{fig:accretion}}
\end{figure}

\begin{figure}
\figurenum{5}
\begin{center}
\includegraphics[width=2.7in,angle=0]{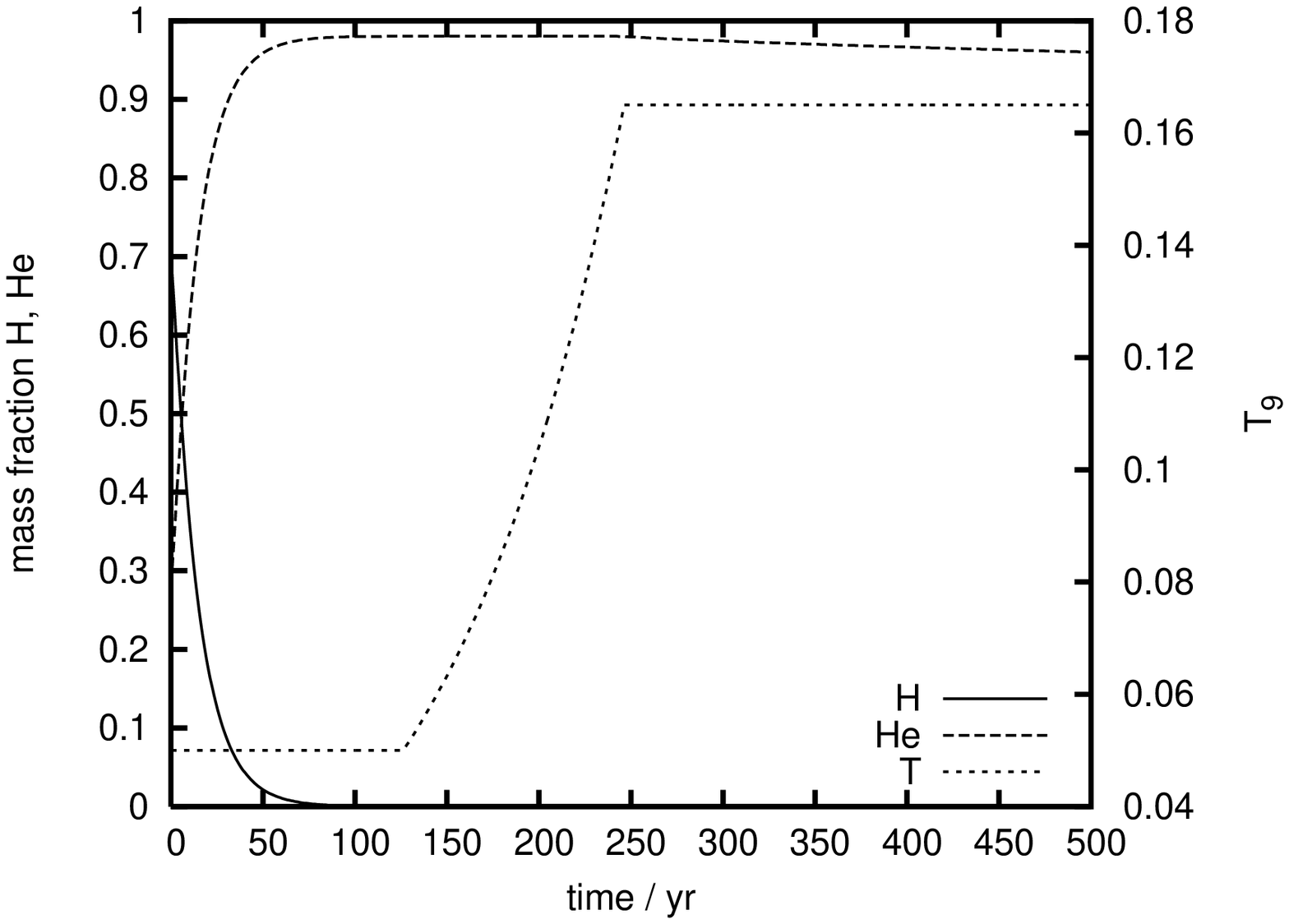}
\includegraphics[width=2.5in,angle=0]{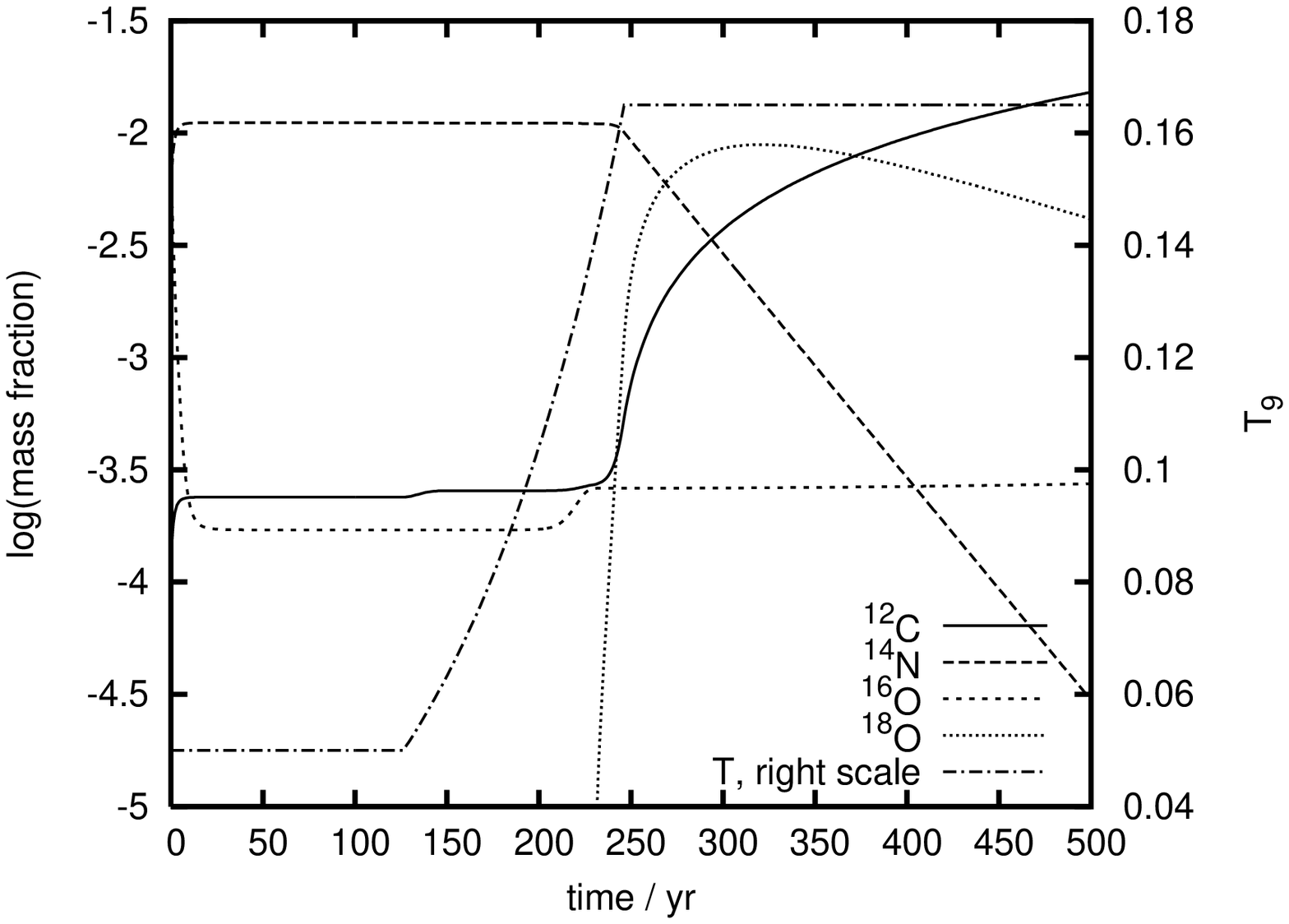}\\
%\vspace{1.5in}
\includegraphics[width=2.35in,angle=0]{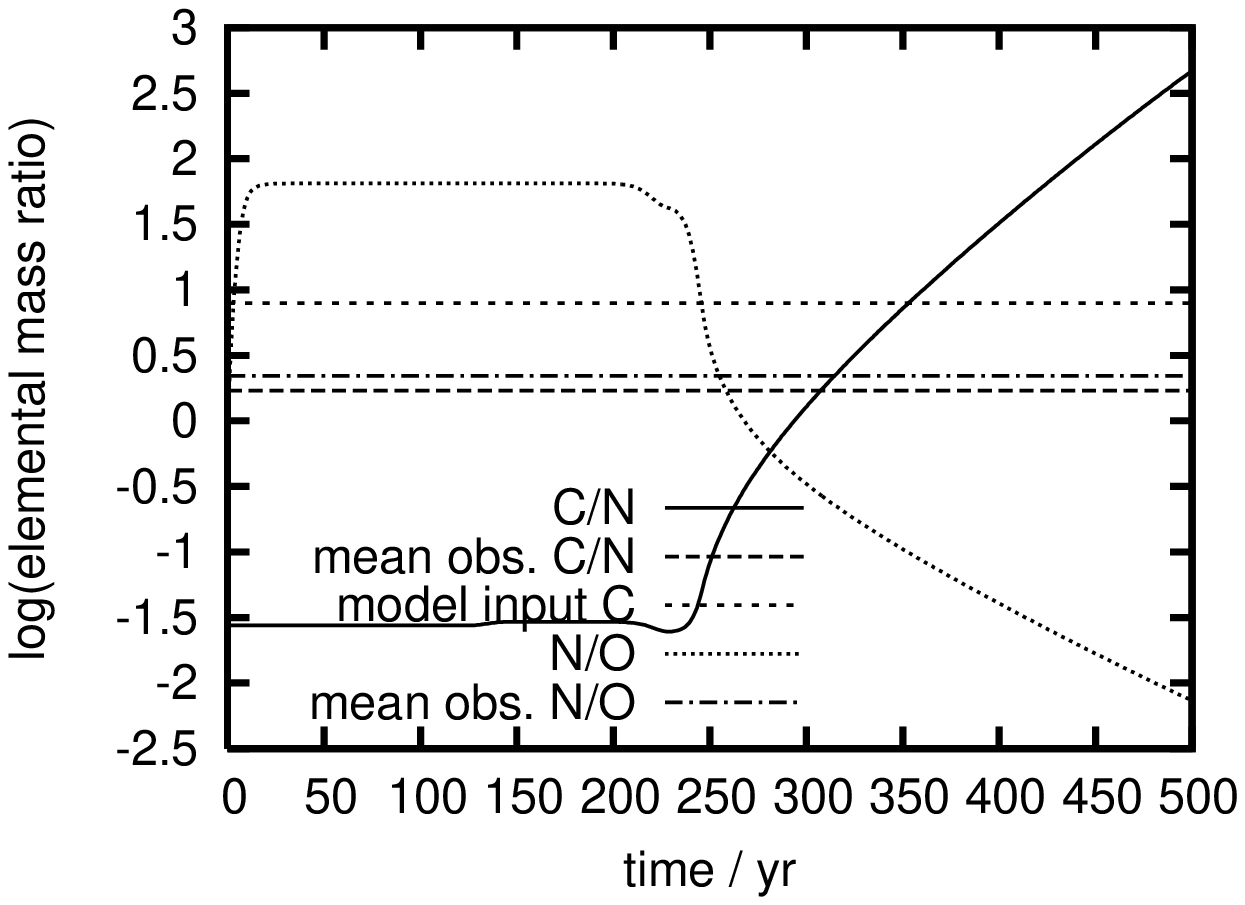}
\hspace{0.3in}
\includegraphics[width=2.35in,angle=0]{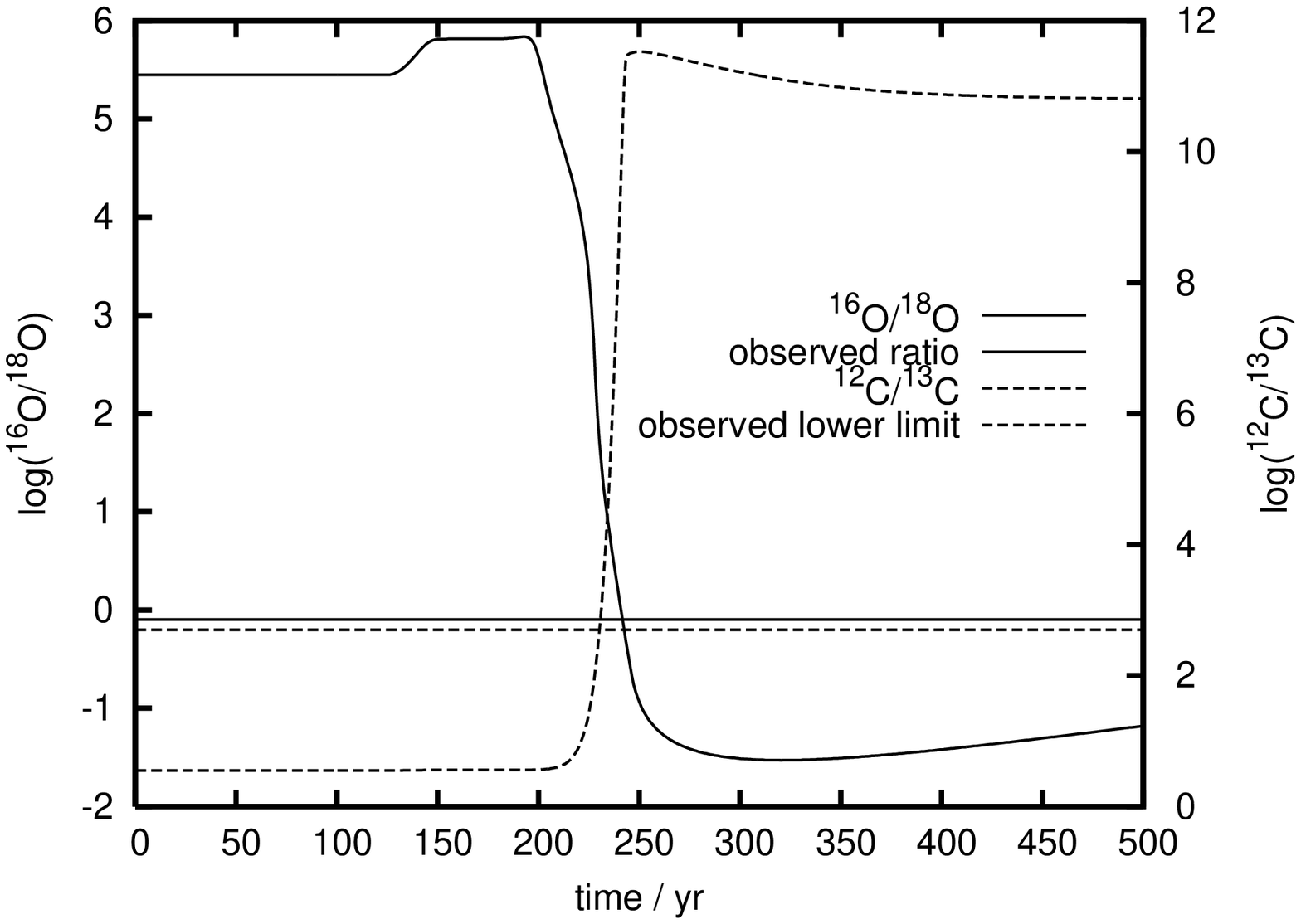}
%\vspace{-2.5in}
\end{center}
\caption{\label{fig:t-H-He-T} Time evolution. Upper left: H and He as well as
temperature. Upper right: CNO isotopes, and temperature. Lower left:
C/N and N/O ratios and mean observed ratios of majority of RCB
stars. Lower right: O and C isotopic ratios and observed ratios
(straight lines). }
\end{figure}

\begin{figure}
\figurenum{6}
\begin{center}
\includegraphics[width=2.5in,angle=0]{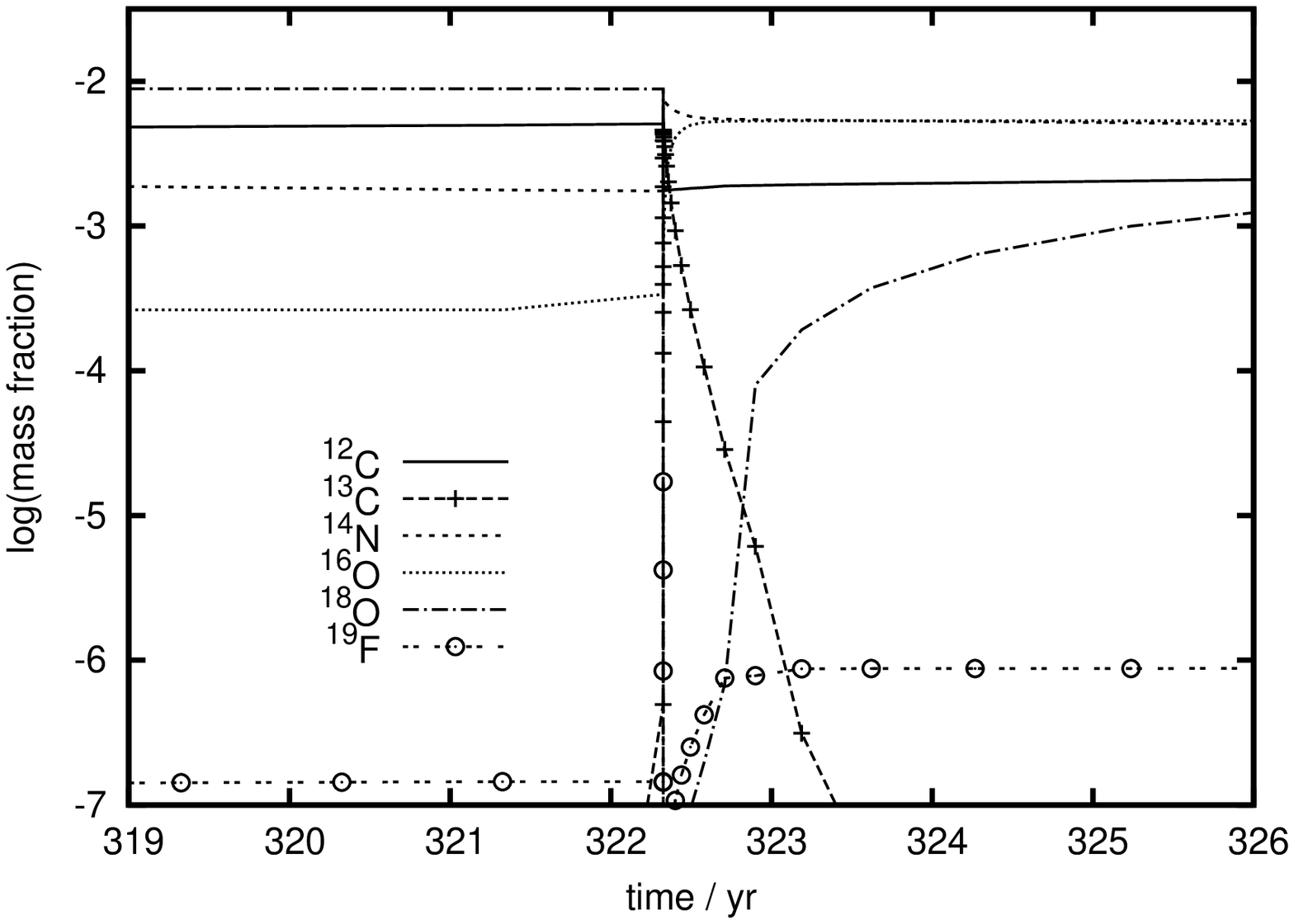}
\includegraphics[width=2.5in,angle=0]{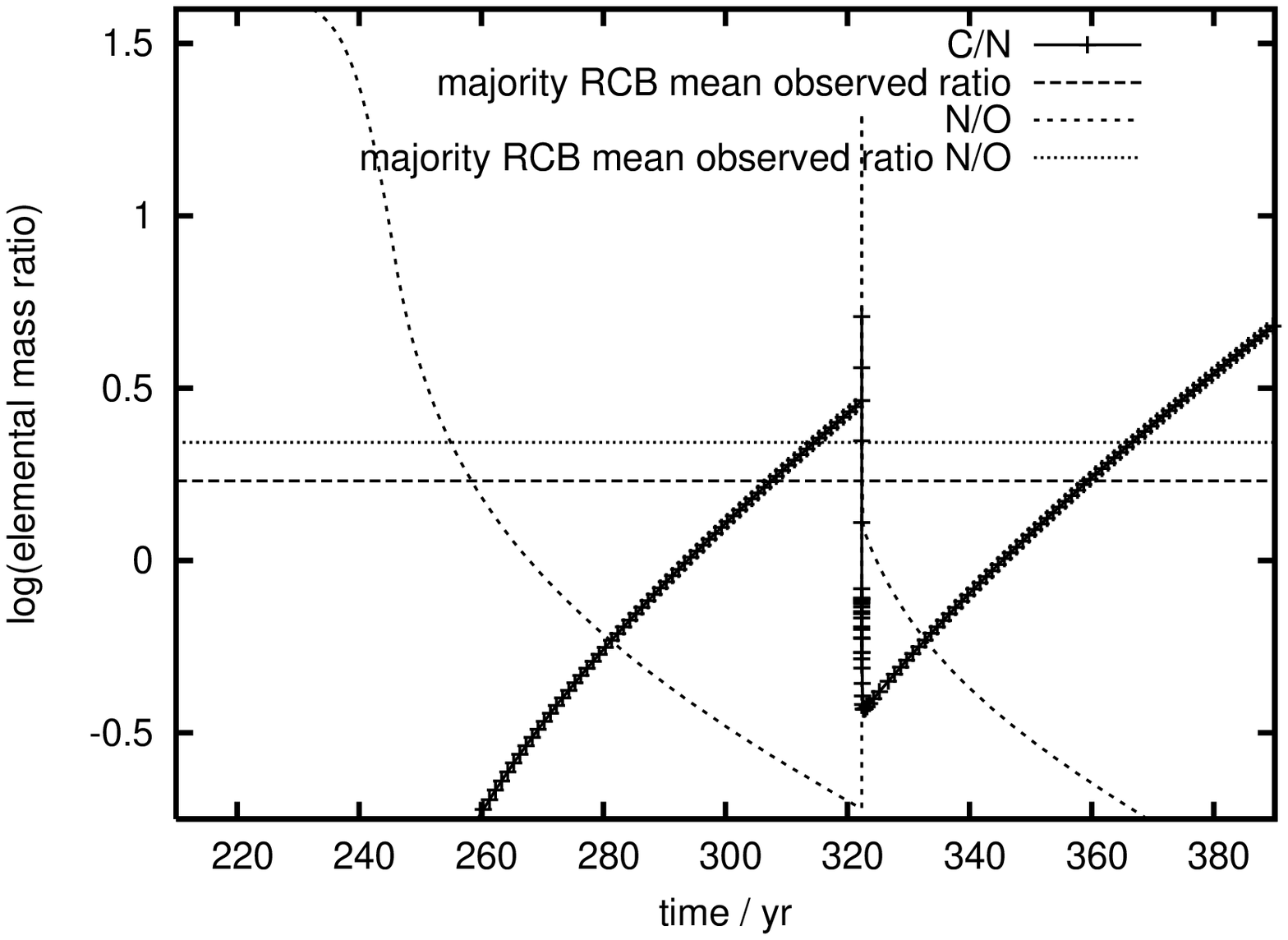}\\
\includegraphics[width=2.5in,angle=0]{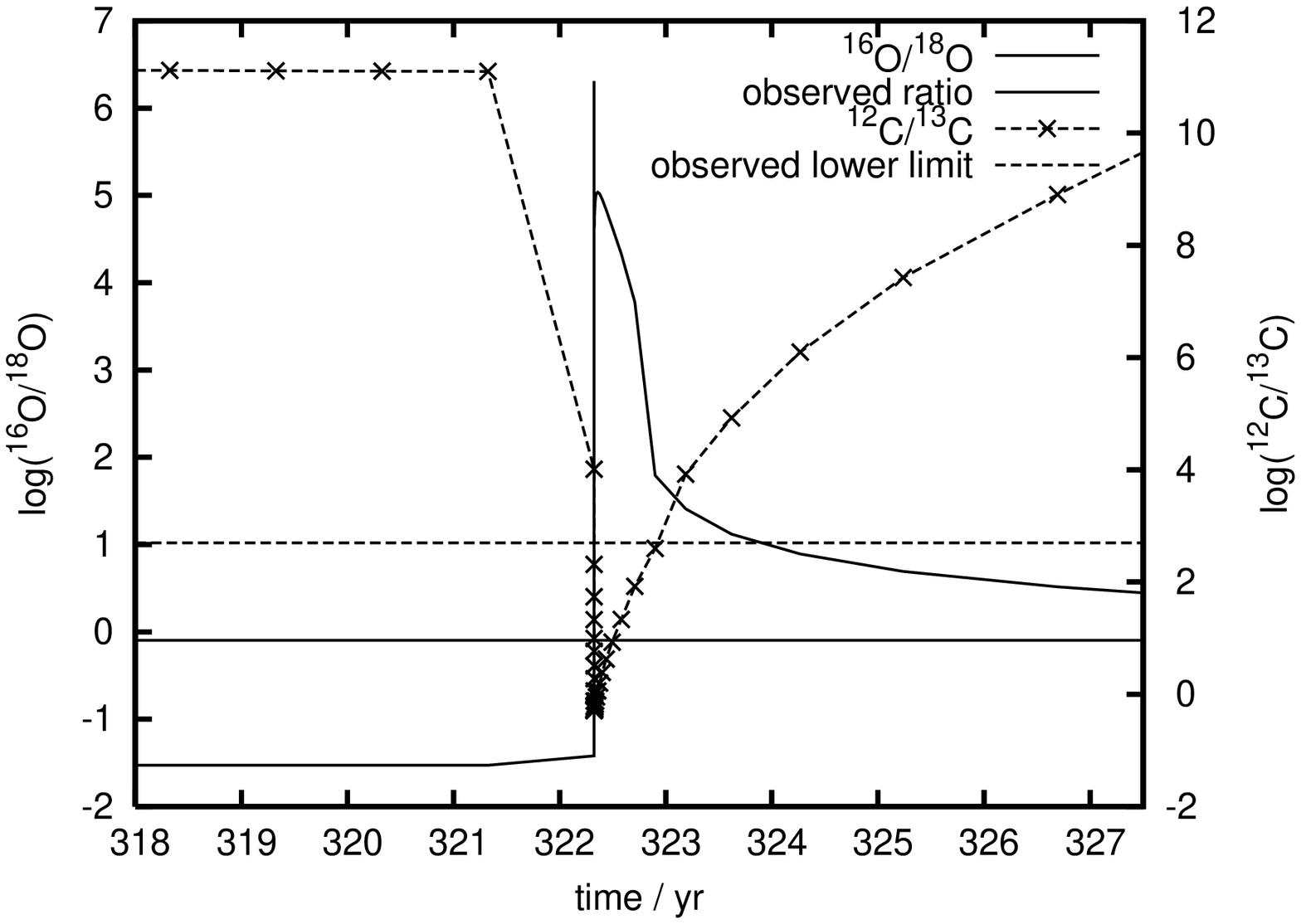}
%\hspace{0.3in}
%\includegraphics[width=2.35in,angle=0]{DDMERGO18_060926/FIGURES/t-OisoCiso-ratios.ps}
\end{center}
\caption{\label{fig:nucl-Hadm}Upper left: Same as \abb{fig:t-H-He-T}
for the short period around the time of admixing the H-rich CO-WD
envelope material to the He-burning accreting material. Time evolution
of CNO isotopes. Each mark along the \cdr\ line represents a time step
in the network evolution. Upper right: Time evolution C/N and N/O
ratios and mean observed ratios of majority of RCB stars. Lower left:
Time evolution of O and C isotopic ratios and observed ratios
(straight lines). }
\end{figure}

%\begin{figure}
%\plotone{DDMERGO18_060926/FIGURES/t-CNO-T_h1.ps} %--> herwig_fig1.eps
%\figcaption{
%Same as \abb{fig:t-CNO-T} for the short period around the time of 
%admixing the H-rich CO-WD envelope material to the He-burning accreting 
%material. Time evolution of CNO isotopes. Each mark along the \cdr\ line 
%represents a time step in the network evolution.
%\label{fig:t-CNO-T_h1}} 
%\end{figure}

%\begin{figure}
%\plotone{DDMERGO18_060926/FIGURES/t-CNOratios_h1.ps} %--> herwig_fig1.eps 
%\figcaption{ Same as \abb{fig:t-CNOratios} for the short period aroundthe time of admixing the H-rich CO-WD envelope material to theHe-burning accreting material. Time evolution C/N and N/O ratios and mean observed ratios of majority of RCB stars.
%\label{fig:t-CNOratios_h1}} 
%\end{figure}

%\begin{figure}
%\plotone{DDMERGO18_060926/FIGURES/t-OisoCiso-ratios_h1.ps} %--> herwig_fig1.eps
%\figcaption{ Same as \abb{fig:t-OisoCiso-ratios} for the short period around the time of admixing the H-rich CO-WD envelope material to the He-burning accreting material. Time evolution of O and C isotopic ratios and observed ratios (straigth lines). 
%\label{fig:t-OisoCiso-ratios_h1}} 
%\end{figure}

\end{document}